\documentclass[aps,prl,reprint,superscriptaddress,showpacs]{revtex4-1} %endfloats 

\usepackage[T1]{fontenc}
\usepackage{graphicx}
\usepackage{amssymb}

\begin{document}

\title{Mechanism for nuclear and electron spin excitation by radio frequency current} 

\author{Stefan Müllegger}
\email[Corresponding author; ]{stefan.muellegger@jku.at}
\affiliation{Institute of Semiconductor and Solid State Physics, Johannes Kepler University Linz, 4040 Linz, Austria.} 

\author{Eva Rauls} 
\email[Corresponding author; ]{eva.rauls@upb.de}
\affiliation{Lehrstuhl f\"{u}r Theoretische Physik, University of Paderborn, 33098 Paderborn, Germany.}

\author{Uwe Gerstmann} 
\affiliation{Lehrstuhl f\"{u}r Theoretische Physik, University of Paderborn, 33098 Paderborn, Germany.}

\author{Stefano Tebi} 
\affiliation{Institute of Semiconductor and Solid State Physics, Johannes Kepler University Linz, 4040 Linz, Austria.} 

\author{Giulia Serrano} 
\affiliation{Institute of Semiconductor and Solid State Physics, Johannes Kepler University Linz, 4040 Linz, Austria.} 

\author{Stefan Wiespointner-Baumgarthuber} 
\affiliation{Institute of Semiconductor and Solid State Physics, Johannes Kepler University Linz, 4040 Linz, Austria.} 

\author{Wolf Gero Schmidt}
\affiliation{Lehrstuhl f\"{u}r Theoretische Physik, University of Paderborn, 33098 Paderborn, Germany.} 

\author{Reinhold Koch} 
\affiliation{Institute of Semiconductor and Solid State Physics, Johannes Kepler University Linz, 4040 Linz, Austria.} 

\newcommand{\didv}{$\mathrm{d}I/\mathrm{d}V$\,}
\newcommand{\tbpc}{TbPc$_2$}

\begin{abstract}
Recent radio frequency scanning tunneling spectroscopy (rf-STS) experiments have demonstrated nuclear and electron spin excitations up to $\pm12\hbar$ in a single molecular spin quantum dot (qudot). 
Despite the profound experimental evidence, the observed independence of the well-established dipole selection rules is not described by existing theory of magnetic resonance -- pointing to a new excitation mechanism. 
Here we solve the puzzle of the underlying mechanism by presenting all relevant mechanistic steps. 
At the heart of the mechanism, periodic transient charging and electric polarization due to the rf-modulated tunneling process cause a periodic asymmetric deformation of the qudot, enabling spin transitions via spin-phonon-like coupling. 
The mechanism has general relevance for a broad variety of different spin qudots exhibiting internal mechanical degrees of freedom (organic molecules, doped semiconductor qudots, nanocrystals, etc.). 
\end{abstract}


%73.50.Rb ... Acoustoelectric and magnetoacoustic effects
%75.78.-n ...	Magnetization dynamics
%75.50.Xx ... Molecular magnets
%68.37.Ef ... Scanning tunneling microscopy (including chemistry induced with STM)

%68.35.Ja ... Surface and interface dynamics and vibrations
%75.76.+j	Spin transport effects

%37.90.+j
%84.40.-x
%33.20.-t 	Molecular spectra
%76.90.+d ... Other topics in magnetic resonances and relaxations
%73.22.-f 	Electronic structure of nanoscale materials and related systems

\maketitle

Recently, we introduced the new technique of radio frequency scanning tunneling spectroscopy (rf-STS) \cite{Mullegger2014d} by investigating a molecular spin quantum dot (qudot), namely a single molecule of the archetypal terbium double-decker (\tbpc) \cite{Ishikawa2005} single-ion magnet (SIM) on a Au(111) substrate. 
By utilizing resonant rf current, we succeeded in demonstrating excitation of electron and nuclear spin transitions in these qudots. 
The observed changes of electron ($J$) and nuclear ($I$) angular momentum components of up to $\Delta J_z=\pm12$ or nuclear $\Delta I_z=\pm3$ \cite{Mullegger2014d} contradict the electromagnetic dipole selection rules ($\Delta J_z=\pm1$ and $\Delta I_z=\pm1$), which govern the well-established methods of electron and nuclear magnetic resonance. 
This points towards a novel spin-excitation channel by rf-STS, which is fundamentally different from photon-induced spin-excitation processes. 
Our rf-STS experiments \cite{Mullegger2014d,Mullegger2014b} revealed that electron tunneling via a molecular orbital (MO) of \tbpc\, is crucial for rf-STS-based spin excitation. 
The necessity of orbital-mediated electron tunneling in rf-STS is in marked contrast to spin excitation by inelastic dc electron tunneling \cite{Heinrich2004,Fu2009,Loth2010}; the latter occurs at characteristic threshold energies symmetrically above and below the Fermi energy; higher spin excitations are sequential two- or three-step processes with $\Delta J_z=\pm1$ for each step \cite{Loth2010}.  
The intriguing evidence for efficient higher spin excitations via one-step excitation processes in rf-STS points towards involvement of mechanical degrees of freedom \cite{Mullegger2014b} for fulfilling the fundamental angular momentum conservation. 

In this letter, we present a mechanism that explains spin excitations by rf-STS in a single molecular spin qudot. 
The mechanism is confirmed by first principles calculations, which show that the internal mechanical degrees of freedom of a molecular spin qudot play a crucial role for spin excitation in an rf-STS experiment. 
At the heart of the mechanism lie the periodic transient charging and electric polarization of the qudot due to the rf-modulated tunneling process. 
The rf modulation of electron tunneling induces periodic electronic and structural perturbations in the qudot -- not present in stochastic dc-STM experiments. 
The resulting mechanical oscillations enable angular momentum transfer between the qudot's spin and its mechanical backbone -- similar to the role of phonons in spin-phonon coupling. 
By lowering the symmetry, thereby relaxing the selection rules, the mechanism becomes very efficient.

\begin{figure}
\includegraphics[width=7.8cm]{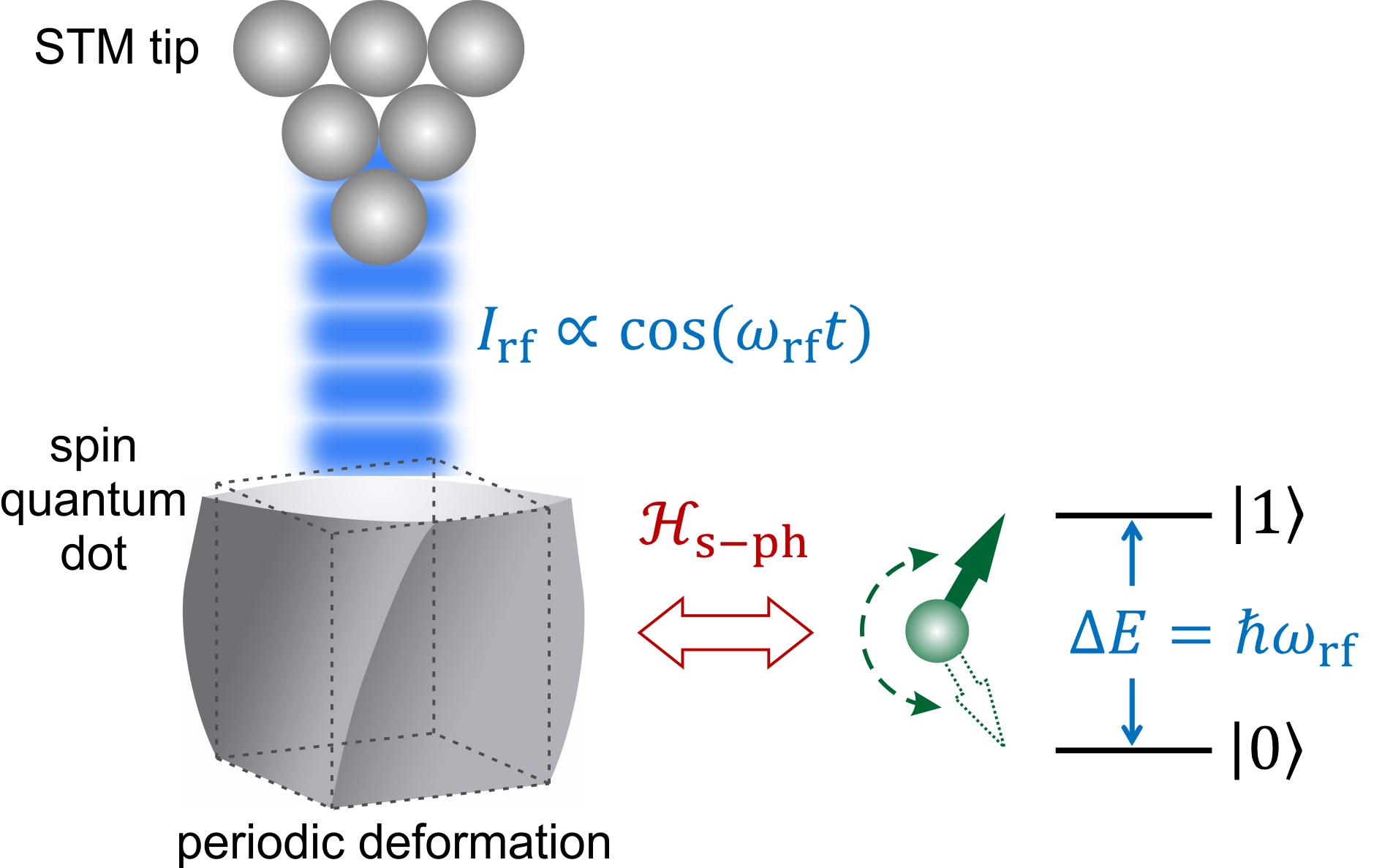}
\caption{\label{fig:scheme} Schematic illustration of resonantly exciting a spin transition in a molecular spin qudot in an rf-STS experiment by mechanical deformation due to rf-modulated tunneling current $I_\mathrm{rf}$.}
\end{figure} 

In an rf-STS experiment (Fig.\,\ref{fig:scheme}), the qudot's mechanical structure is periodically deformed by the rf-modulated electron tunneling process at frequency $\omega_\mathrm{rf}$ \cite{Mullegger2014b}. 
The periodic modulation adds a periodic component to the initially random tunneling of electrons. 
Based on the experimental results of Ref.~\onlinecite{Mullegger2014d}, a comprehensive description of the rf-STS mechanism has to take into account the transient electric charging of the qudot, electric ($E$) field- as well as STM tip and substrate effects. 
Charging and $E$-field can affect simultaneously the mechanical structure of the molecule during an rf-STS experiment. 
To investigate details of the mechanism by theoretical modeling, the simulation of a model system as realistic as possible, i.\,e. including a proper description of the substrate, is mandatory. 
Herein, the model system is represented by single \tbpc\, molecules on Au(111), as studied in Ref.~\onlinecite{Mullegger2014d}. 
\tbpc\, consists of a Tb$^{3+}$ ion with total electronic angular momentum of $J=6$ and nuclear spin $I=3/2$ sandwiched between two organic phthalocyanine (Pc) ligands \cite{Ishikawa2005}. 
The ligand field stabilizes a strong uniaxial (i.\,e., Ising-like) magnetic anisotropy in this SIM with a ground-state doublet $J_z=\pm6$ \cite{Thiele2013}. 

\begin{figure}
\includegraphics[width=6cm]{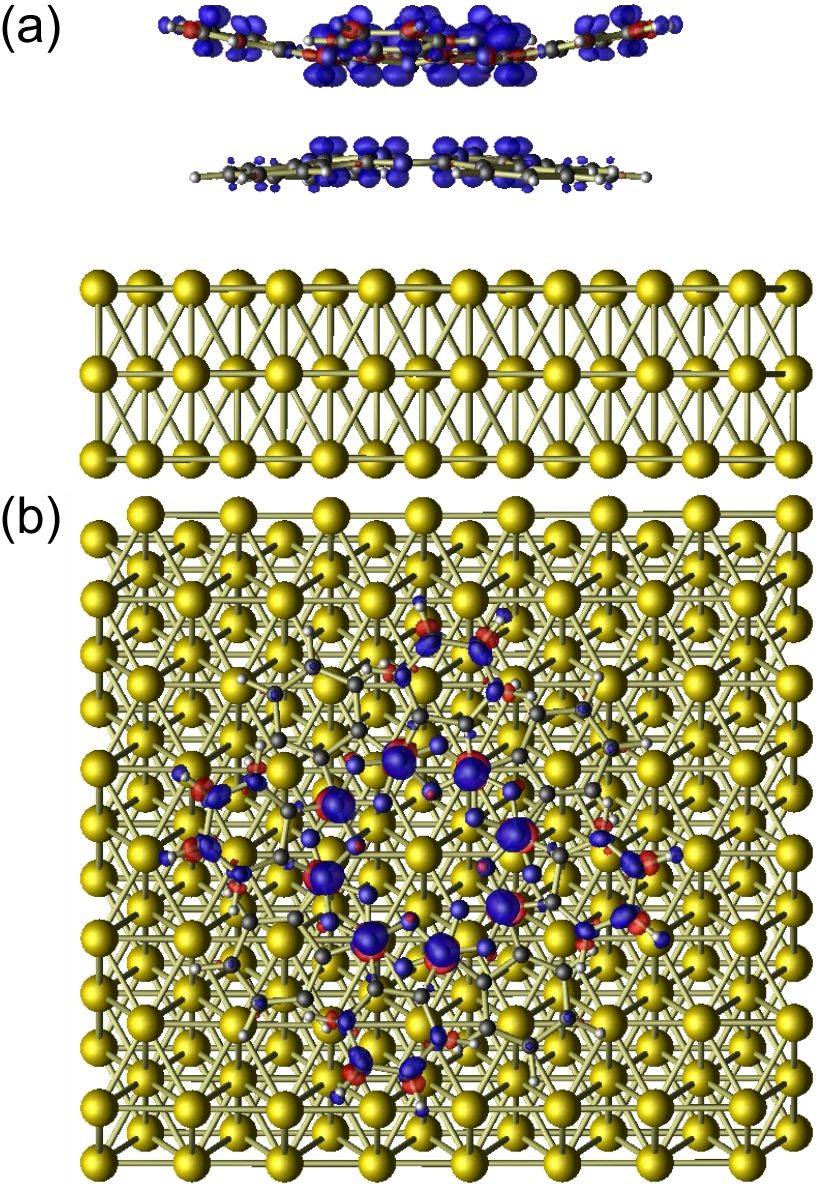}
\caption{\label{fig:neutral} Side (a) and top (b)  view of the charge density of the LUMO (occupied part of degenerate LUMOs) of the unperturbed neutral [\tbpc]$^0$ molecule adsorbed on Au(111).}
\end{figure}

As shown in Ref.~\onlinecite{Mullegger2014d}, the observation of an rf-STS signal relies on electron flow through an MO of the molecular qudot. 
Each tunneling electron transiently charges the molecular qudot. 
If the electron lifetime in the MO (before leaving to the substrate) $\tau_\mathrm{MO}$ is sufficiently long, the transient electron polarizes the qudot, slightly altering atom positions and bond lengths within the dot by partial de-/re-population of MOs. 
The situation is similar to the temporary formation of a molecular polaron (i.\,e. the quantum of lattice deformation caused by electric polarization by one electron). 
For \tbpc-based tiny qudots, the lowest unoccupied MO (LUMO), utilized in Ref.~\cite{Mullegger2014d}, is essentially a $\pi$ orbital of the Pc ligands \cite{Takamatsu2007}. 
From the lifetime of excited charge states in a Pc monolayer on metal surface a value of $\tau_\mathrm{MO} \approx 10^{-9}$~s was estimated by Takaki et al.~\cite{Takagi2002}, which is a value similar to that of the bulk phase \cite{Ueno2012}. 
In contrast, the mechanical structure of the molecular backbone is adopted at a much shorter timescale of $10^{-12}$~s, assuring (quasi) equilibrium conditions for the charge-induced structural change. 
The transiently charged qudot adopts a ground state configuration with a different mechanical structure compared to the neutral one \cite{Takamatsu2007}. 
This polarization induced mechanical deformation is similar to the polaron-induced mechanical switching of single-molecule junctions discussed by Galperin et al. \cite{Galperin2005}. 

For \tbpc/Au(111), we study herein the polaron effect upon transient charging with first principles calculations. 
For this purpose we add one extra electron into the LUMO of the neutral structure. 
For the DFT calculations we employ the Vienna Ab Initio Simulation Package \cite{a-vasp} including a supercell approach and periodic boundary conditions for an accurate description of the electronic structure of the metallic Au(111) surface. 
The latter is crucial for a realistic occupation of the molecular orbitals of the molecular qudot. 
Electron/ion interactions are described by the projector-augmented wave method \cite{c-paw} and the electron-electron exchange-correlation energy by the generalized gradient approximation \cite{b-gga}. 
The influence of dispersive interactions is included via the semi-empirical London dispersion formula \cite{d-ortmann}. 
The $4f$-shell of the Tb ion is fixed in a trivalent configuration, in agreement with the Tb$^{3+}$-like spectra experimentally observed in Ref.~\onlinecite{Mullegger2014d}. 
We use an energy cut-off of 400~eV for the plane waves and the $\Gamma$--point for the Brillouin zone modeling. 
The Au-surface is modeled by three layers, the lowest kept fixed at bulk position during relaxation, while all other atoms are allowed to relax freely. 
A vacuum layer of 2.5~nm is found to be sufficient to avoid artificial interactions between periodic images parallel to the substrate normal ($z$-axis). 
As in the gas phase, the LUMO of neutral [\tbpc]$^0$ adsorbed on Au(111) consists of a superposition of energetically degenerate orbitals (Fig.~\ref{fig:neutral}). 
Hence, without external perturbation a symmetric distribution of the extra electron is induced, predominantly localized at the upper Pc ligand. 

\begin{figure} 
\includegraphics[width=6.5cm]{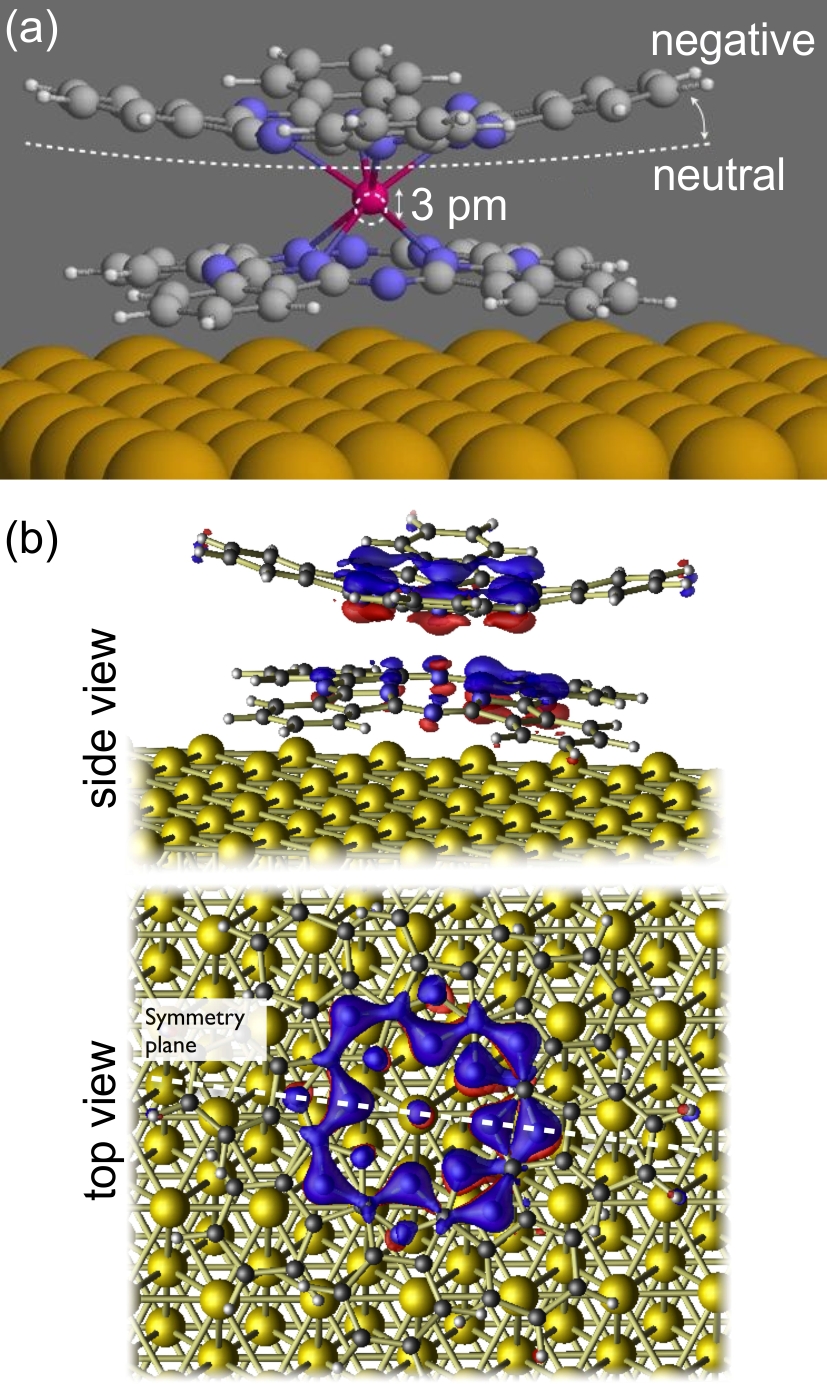}
\caption{\label{fig:charge} Negatively charged [\tbpc]$^-$ ion on Au(111). (a) DFT structure; dotted white lines indicate the structure of the neutral [\tbpc]$^0$ molecule. (b) Distribution of the extra electron in the charged state after full electronic and structural relaxation; blue (red) indicates electron accumulation (depletion).}
\end{figure}

In a first step, as a starting model for the transient charging process, we add a permanent extra electron to the system, i.\,e. we investigate [\tbpc]$^-$ on Au(111). 
Figure~\ref{fig:charge}a illustrates the resulting polaronic deformation.
Compared to the neutral form (dotted lines), negatively charged [\tbpc]$^{-}$ exhibits a more strongly bent upper Pc ligand and increased vertical distance to the substrate surface. 
In particular, the Tb$^{3+}$ ion lies $\approx$3~pm farther away from the substrate. 
In this respect, our results are consistent with earlier gas phase calculations by Takamatsu et al. \cite{Takamatsu2007} reporting that neutral [\tbpc]$^0$ has a 10~pm shorter Pc-Pc distance than [\tbpc]$^-$ due to increased bond order by emptying an antibonding MO. 
However, to understand further details of the mechanism, the substrate and the position of the tip have to be taken into account. 
Due to the substrate-induced metallic occupation of the MOs, very small non-centrosymmetric perturbations are sufficient to lift the degeneracy of the LUMO orbitals, giving rise to some structural relaxation as well as an asymmetric charge density, both further increasing the mechanical angular momentum. 

Figure~\ref{fig:charge}b shows that after full structural relaxation, the induced charge distribution in [\tbpc]$^-$ upon tunneling into the LUMO becomes inhomogeneous: asymmetric in charge and also slightly in structure (due to weak symmetry breaking by the substrate). 
The pyrrole groups of the upper molecule are no longer at the same height above the surface but differ by $\approx$2~pm, whereby the phtalocyanine ligand with the larger part of the induced density of the extra electron is stronger bent and lifted. 
Slightly smaller is the difference for the nitrogen atoms.  
It is reasonable to assume that this plot obtained for the negatively charged species actually represents a snapshot of the transient charging process just before the electron in the LUMO flows off to the substrate; remember, it stays about 1~ns there. 
It is important to note here, that small external perturbations such as, e.\,g., a small off-center positioning of the STM tip are expected as further efficient driving forces to break the symmetry, as well.

\begin{figure}
\includegraphics[width=5.8cm]{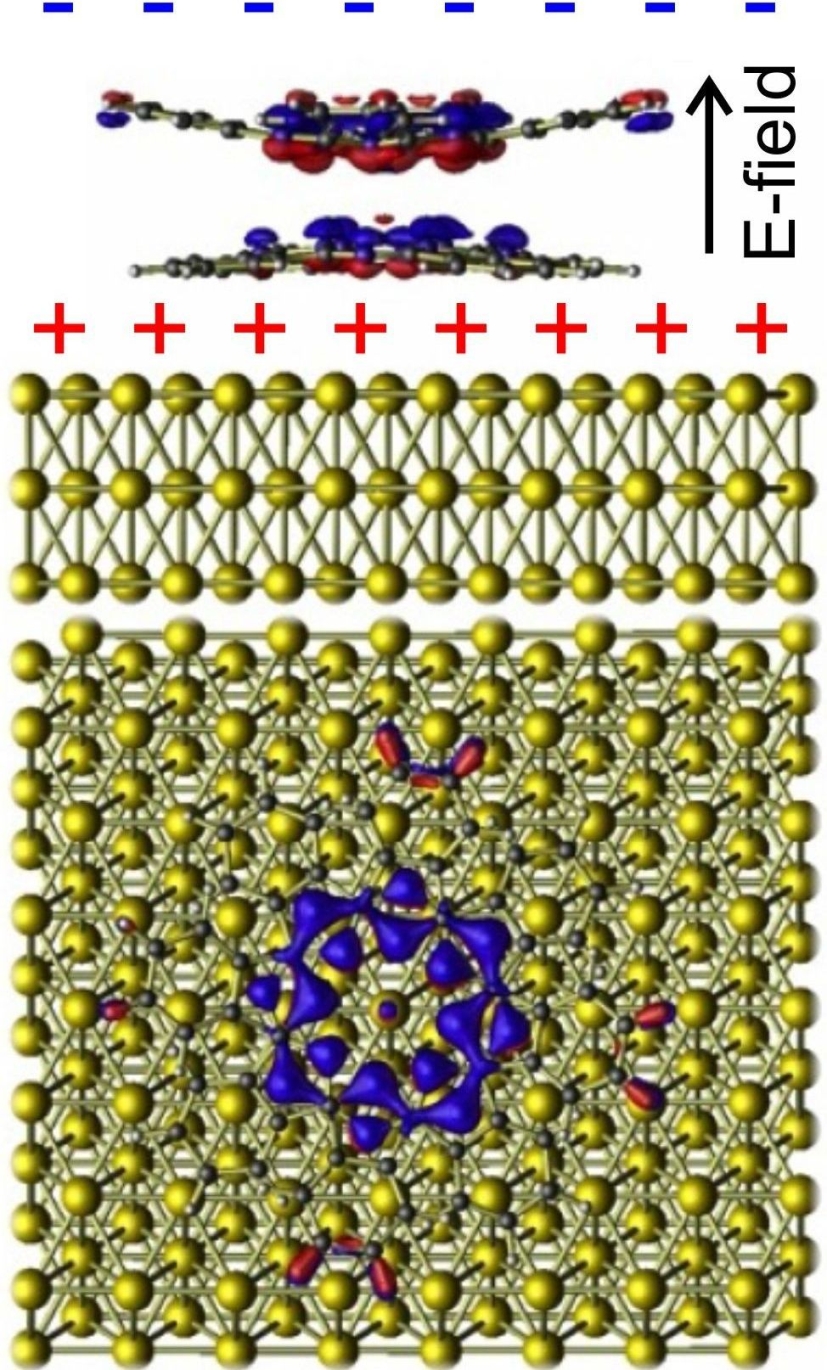}
\caption{\label{fig:efield} DFT relaxed structure of neutral [\tbpc]$^0$ on Au(111) under the presence of an electric field of 5~V/nm perpendicular to the surface; blue (red) indicates electron accumulation (depletion).}
\end{figure}

The relaxation towards an asymmetric configuration may also be facilitated by the $E$-field due to the bias voltage between sample and STM tip.  
To see that, it is illustrative to discuss the effect of the $E$-field separately, i.\,e. independent of the above mentioned charging mechanism. 
Hence, onto the neutral system we have applied a homogeneous $E$-field~\cite{makov95} perpendicular to the substrate. 
Like in experiment, the system behaves critically with respect to the $E$-field strength: Visible effects are obtained in a range of 0.5\,--\,5~V/nm. 
Above 5~V/nm the \tbpc\, molecule starts to dissociate. 
This is in full agreement with our experimental $E$-field value of 4~V/nm causing dissociation of \tbpc\, molecules on Au(111). 

Figure~\ref{fig:efield} shows that an $E$-field of 5~V/nm induces a structural relaxation of neutral [\tbpc]$^0$ similar to that of the negatively charged [\tbpc]$^-$ species (shown in Fig.~\ref{fig:charge}b). 
Here, the structural relaxation is caused by a shift of the energy levels of the \tbpc\, molecule with respect to the Fermi level, resulting in a redistribution of electron density. 
The applied $E$-field increases the \mbox{Pc-Pc} distance by nearly 30~pm, whereby the Tb ion lies $\approx16$~pm farther away from the substrate. 
Similar to a rigid charging, the presence of an $E$-field induces electron density in anti-bonding MOs representing the LUMO of the neutral [\tbpc]$^0$ species. 
This effect becomes also visible by the corresponding charge redistribution (Fig.~\ref{fig:efield}) which strongly resembles that of the extra electron in the negatively charged system (Fig.~\ref{fig:charge}b), in particular reflecting the same reduced symmetry. 
Notice, that the rf-STS experiments of Ref.~\onlinecite{Mullegger2014d} were performed at $E\approx1.1$~V/nm, where the $E$-field effect is considerably smaller. 

Summarizing our DFT results, the simulations confirm that neutral and charged \tbpc\, molecules have different mechanical structures. 
Our findings reveal that the interactions of \tbpc\, with the substrate and/or the STM tip slightly lift the degeneracy of its frontier MOs. 
Such small perturbations are necessary and sufficient for an asymmetric density distribution of the anti-bonding LUMO of \tbpc. 
This asymmetry causes asymmetric mechanical deformation of \tbpc, giving rise to nonzero shear components in the respective strain tensor, crucial for the rf-STS mechanism.
Due to the reduced symmetry, selection rules are relaxed enabling higher spin transition, as observed in Ref.~\onlinecite{Mullegger2014d}. 
By this, our calculations provide a natural explanation for the rf-STS-induced spin excitation:  
From magnetic resonance on bulk samples it is well-known that angular momentum may readily be transferred between spin- and mechanical degrees of freedom. 
The latter are described by the quasiparticle-picture of phonons. 
Spin-phonon effects play an important role, e.\,g., for conservation of angular momentum in the famous Barnett- \cite{Barnett1915} and Einstein de Haas effect \cite{EinsteindeHaas1915} and the decoherence of spin polarization. 
Moreover, angular momentum transfer pathways have been experimentally observed between spins and thermal phonons \cite{Tikhonov2013}, ultrasonic waves in molecular crystals \cite{Calero2007}, and mechanical modes of nanoresonators \cite{Garanin2011,Chudnovsky2014}. 
In molecular magnets, spin Rabi-oscillations excited by ultrasound in the GHz regime were found \cite{Calero2007b}. 
Similar to the case of spin-phonon coupling, here, the periodic mechanical deformation of \tbpc\, is responsible for an additional (perturbative) component in the spin Hamiltonian. 
The latter depends on the (periodically in time modulated) positions of \tbpc's atoms, thus establishing a coupling of mechanical- with spin degrees of freedom. 
The additional periodic perturbation of the spin Hamiltonian couples different spin states of \tbpc. 
Eigenstates of the unperturbed system are no longer eigenstates of the perturbed one, which is described by a superposition of different spin states of \tbpc. 
This mixing of states enables transitions between different spin states, such as periodically excitated spin transitions at resonance, i.\,e. for $\omega_\mathrm{rf}=\Delta E_\mathrm{spin}/\hbar$ as reported in Ref.~\onlinecite{Mullegger2014d} (Fig.~\ref{fig:scheme}). 

In conclusion, the spin excitation mechanism presented herein was introduced by means of spin qudots represented by single terbium (III) double-decker molecules, which exhibit a rich spectrum of mechanical degrees of freedom.  
We remark, however, that the presented mechanism for rf-induced spin excitation has a much more general relevance. 
It can be extended generally to a broad variety of different spin qudots exhibiting internal mechanical degrees of freedom (organic molecules, doped semiconductor qudots, nanocrystals, etc.). 
Furthermore, the proposed mechanism is not restricted to tunneling electrons but valid also for other types of conductance. 

\begin{acknowledgments}
We kindly acknowledge financial support of the project I958 by the Austrian Science Fund (FWF) and the Deutsche Forschungsgemeinschaft (DFG) as well as the DFG priority program SPP~1601. 
\end{acknowledgments}

%\bibliography{TbPc2}


\end{document}